\documentclass[a4paper]{jpconf}
\usepackage{amssymb}
\usepackage{graphicx}

\newcommand{\be}{\begin{equation}}
\newcommand{\ee}{\end{equation}}
\newcommand{\bra}{\langle}
\newcommand{\ket}{\rangle}
\newcommand{\bea}{\begin{eqnarray}}
\newcommand{\eea}{\end{eqnarray}}

\begin{document}
\title{Dynamical Analysis of Stock Market Instability by Cross-correlation Matrix}

\author{Tetsuya Takaishi}

\address{Hiroshima University of Economics, Hiroshima 731-0192, JAPAN}

\ead{tt-taka@hue.ac.jp}

\begin{abstract}
We study stock market instability by using cross-correlations constructed from the return time series of 366 stocks traded on the Tokyo Stock Exchange from January 5, 1998 to December 30, 2013.
To investigate the dynamical evolution of the cross-correlations, cross-correlation matrices are calculated with a rolling window of 400 days.
To quantify the volatile market stages where the potential risk is high, we apply the principal components analysis and measure the cumulative risk fraction (CRF), which is the system variance associated with the first few principal components.
From the CRF, we detected three volatile market stages corresponding to the bankruptcy of Lehman Brothers, the 2011 Tohoku Region Pacific Coast Earthquake, and the FRB QE3 reduction observation in the study period.
We further apply the random matrix theory for the risk analysis and find that the first eigenvector is more equally de-localized when the market is volatile.
\end{abstract}

\section{Introduction}
The stock market is a complex system that
undergoes unstable periods that result in financial crises in some cases.
Measuring systemic risk is an important task to monitor current market status and possibly to avoid
a future financial crisis.
In financial crises, many stocks are interconnected and move collectively.
The level of interconnectedness can be measured by cross-correlations between stocks, and
there are a variety of works on cross-correlations that include the random matrix theory (RMT)\cite{RMT1,RMT2,RMT3,RMT4,RMT5}
and the principal component analysis (PCA)\cite{SR1,SR2,SR3,SR4}.
In this study, we calculate cross-correlations between stocks traded on the Tokyo Stock Exchange from January 5, 1998 to December 30, 2013 and
apply the PCA and the RMT to analyze the dynamical properties of cross-correlations.
In particular, we focus on the market instability  and investigate when the market was volatile during the study period. 

\section{Cross-correlation matrix}

We analyze the daily closing price data of stocks
traded on the Tokyo Stock Exchange from January 5, 1998 to December 30, 2013, which corresponds to 3932 working days.
We choose 366 stocks listed on the Topix 500 index.

\begin{figure}
\centering
\includegraphics[height=5.2cm]{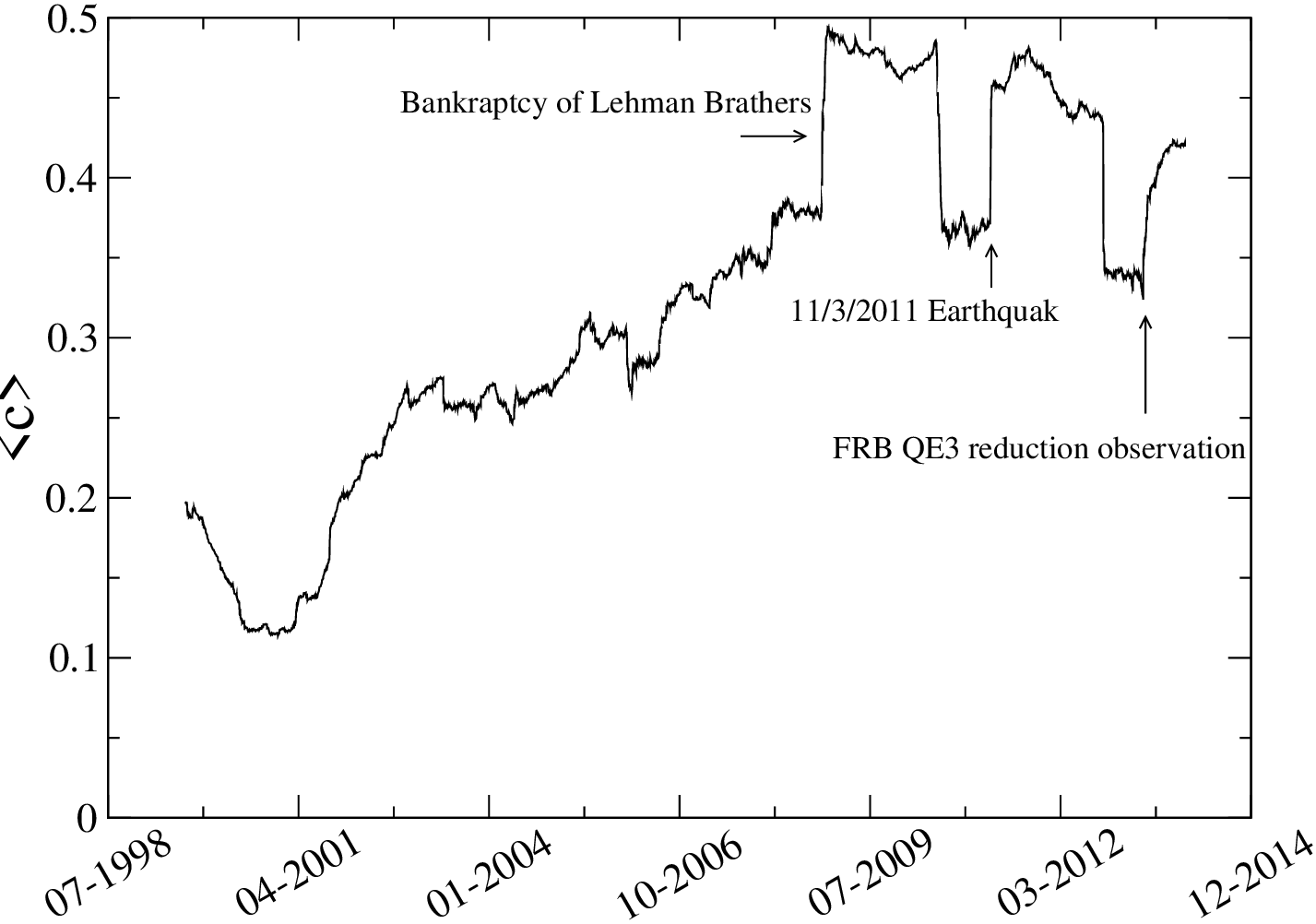}
\hspace{2mm}
\includegraphics[height=5.2cm]{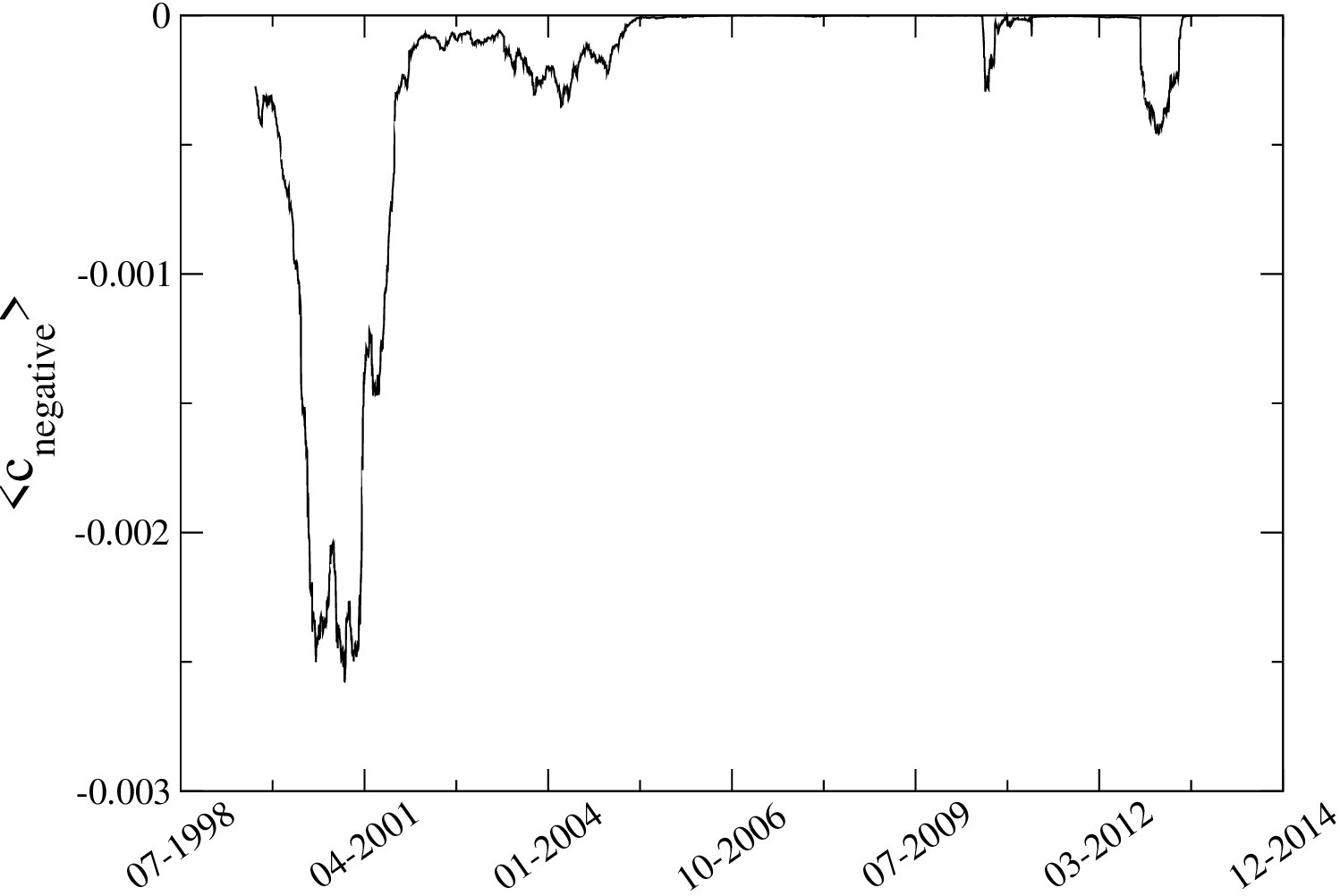}
\caption{
(Left) Average off-diagonal elements of the cross-correlation matrix. (Right) Average negative off-diagonal elements of the cross-correlation matrix.
Each average is taken over 400 days in the rolling window.
}
\vspace{-2mm}
\end{figure}

Let $r_i(t)$ be a return for stock $i$ $(i=1,...,N)$ at time $t$ $(t=1,...,T)$ defined by
the log-price difference as
\be
r_i(t)=\ln p_i(t) -\ln p_i(t-1),
\ee
where $p_i(t)$ is the price for stock $i$ on day $t$.
We also define the normalized return $m_i(t)$ as
\be
m_i(t)=\frac{r_i(t)-\bra r_i \ket}{\sigma_i},
\ee
where $\bra ... \ket$ indicates the time series average and
$\sigma_i$ is the standard deviation of $r_i$.

Using the normalized return $m_i(t)$,
an equal-time cross-correlation matrix is calculated as $c_{ij} =\bra m_i m_j \ket$,
where an average, i.e. $\bra ... \ket$, is taken over a period of the rolling window.
In this study, we consider a rolling window of 400 working days, which roughly corresponds to two years.
By definition, the elements of the cross-correlation matrix are restricted to $-1 \le c_{ij} \le 1$.

Fig.1(Left) shows the dynamical evolution of the average off-diagonal matrix element $\bra c \ket$ given by
\be
\bra c \ket =\frac2{N(N-1)} \sum_{i>j}^N c_{ij},
\ee
where $N=366$.
From the figure, we recognize that there exist three points where $\bra c \ket$ increases abruptly.
According to the historically observed events, these points correspond to the bankruptcy of Lehman Brothers,
the Tohoku region pacific coast earthquake on 11/3/2011, and the FRB QE3 reduction observation
as indicated in the figure.
In Fig.1(Right), we also show the average of negative off-diagonal elements,
\be
\bra c_{negative} \ket =\frac2{N(N-1)} \sum_{i>j, c_{ij}<0}^N c_{ij}.
\ee
Notably, in the recent years, the contribution of negative off-diagonal elements to the cross-correlation matrix
becomes less than that around 2000. 
In particular, at volatile stages, negative off-diagonal components disappear and most stocks are
positively correlated.

\begin{figure}
\centering
\includegraphics[height=5.3cm]{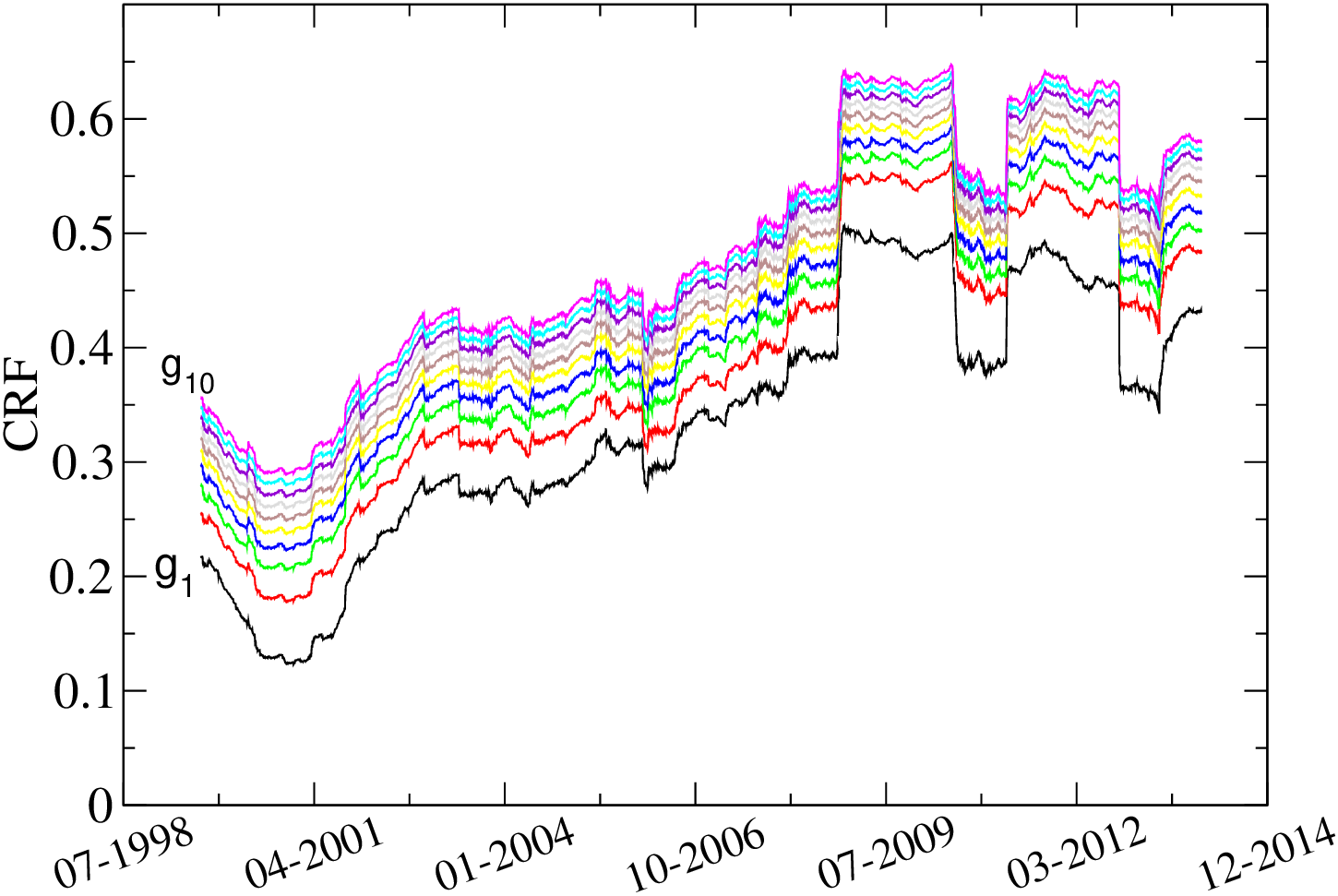}
\includegraphics[height=5.3cm]{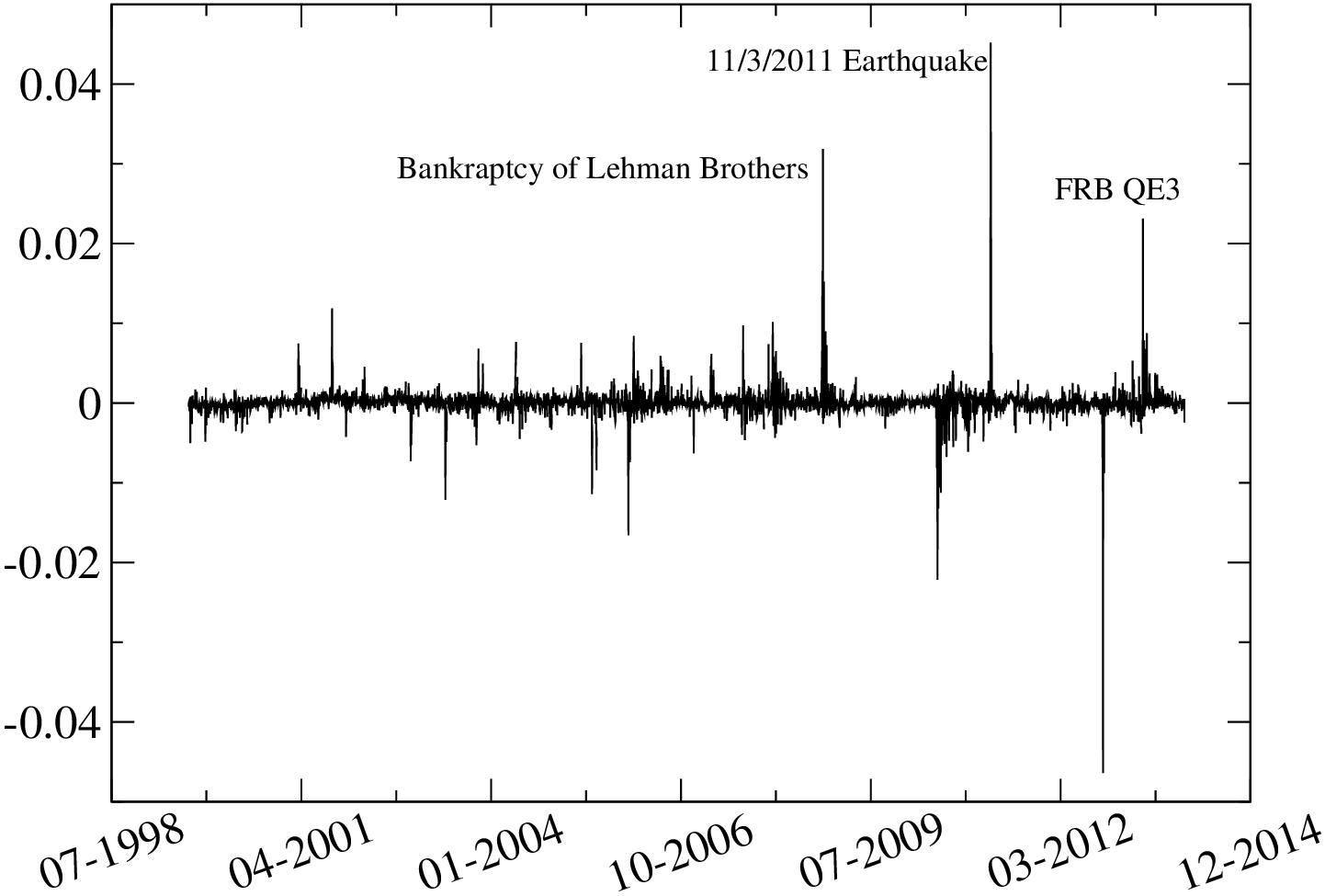}
\caption{
(Left) Time evolution of the CRF with 400-day rolling window
and (Right) the change of $g_1$.
}
\vspace{-2mm}
\end{figure}

\section{Dynamical behavior of Cumulative Risk Fraction}

In order to further investigate the dynamical properties of cross-correlation matrices,
we apply the principal component analysis (PCA).
Billio {\it et al.}\cite{SR1} suggested to use the PCA to quantify
the systemic risk and introduced the cumulative risk fraction (CRF) as a risk measure.
The PCA has also been used to measure the systemic risk\cite{SR2,SR3,SR4}.
To construct the CRF,
we first compute the eigenvalues of cross-correlation matrices, denoted as $\lambda_1,\lambda_2,\dots,\lambda_N$, where
all eigenvalues are sorted as
$\lambda_1>\lambda_2>\dots>\lambda_N$.
Then, we calculate the CRF defined by\cite{SR1}:
\be
g_k= \frac{\omega_k}{\Omega},
\ee
where
$\Omega$ is the total variance of the system given by
$\Omega= \sum_{i=1}^N \lambda_i$ and
$\omega_k$ is the risk associated with the first $k$ principal components
given by $\omega_k= \sum_{i=1}^k \lambda_i$.
The CRF quantifies the portion of the total variance  explained by the first $k$ principal components
over the total variance\cite{SR2}.
Usually, the first few principal components explain most of the system variance.
In the periods of financial crisis, many stocks are highly interconnected and their prices easily move together.
In such periods, the CRF is expected to increase considerably because the system variance also increases.

Fig.2(Left) shows the time evolution of the CRF $g_k$ for $k=1,\dots,10$.
We find that the structures of time evolution of $g_k$ for $k=1,\dots,10$ are very similar.
This indicates that the first eigenvalue dominates in the CRF.
We also find that the structure of the CRF resembles that of average off-diagonal elements of the cross-correlation matrix,
and the CRF increases abruptly at the same points as observed in the CRF, that is, Fig.1(Left).

\section{Changes of the Cumulative Risk Fraction}
Zheng {\it et al.}\cite{SR3} introduced the changes of the CRF to
effectively quantify the points where the potential risk is high.
The changes of the CRF is defined by
\be
change_k(t)=g_k(t+1)-g_k(t).
\label{Change}
\ee
The time evolution of $change_1(t)$ is presented in Fig.2(Right).
We find that the change of the CRF shows pronounced positive spikes at the same three points
where we observed the three abrupt increases in the CRF.
Note that large negative spikes are artificially caused by the period of the rolling window.

\section{Random Matrix Theory}
Let $y_i(t)$ be an independent, identically distributed random variable with $i=1,...,N$ at time $t=1,...,T$.
Then, we define the normalized variable:
\be
w_i(t)=\frac{y_i(t)-\bra y_i \ket}{\sigma_{y_i}},
\ee
where $\sigma_{y_i}$ is the standard deviation of $y_i$.
The equal time cross-correlation between variables $y_i(t)$
is given by $ W_{ij}=\bra w_i w_j \ket$.
The matrix $W$ is called Wishart matrix.
For $N\rightarrow \infty$ and $T\rightarrow \infty$ with $Q=T/N>1$, an eigenvalue distribution of the matrix $W$ is theoretically given by\cite{W1,W2}:
\be
\rho(\lambda)=\frac{Q}{2\pi}\frac{\sqrt{(\lambda_+ -\lambda)(\lambda-\lambda_- )}}{\lambda}, \hspace{3mm}
\lambda_\pm =1+\frac1Q\pm 2\sqrt{\frac1Q}.
\ee
Fig.3(Left) compares an eigenvalue distribution of the matrix $W$ with that of the empirical cross-correlation matrix $c$,
where $Q=400/366\approx 1.1$.
The eigenvalue distribution of the empirical cross-correlation matrix differs considerably from the result from the RMT expectation.
In particular, we find that for the empirical cross-correlation matrix, there exist many eigenvalues less than $\lambda_-$ and larger than $\lambda_+$.

Another interesting entity, the inverse partition ratio that characterizes the eigenvectors, is defined by
\be
IPR_k =\sum_{j=1,}^N (v_k^j)^4,
\ee
where $v_k^j$ is the j-th component of the eigenvector for the k-th eigenvalue.
In the RMT, the eigenvector components are de-localized and distributed as a Gaussian distribution.
In such a case, the expectation of the IPR is $3/N$. On the other hand, when the eigenvector components are localized, for example, only one component has a non-zero value,
the expectation of the IPR would be  1.
There also exists another de-localized case in which all eigenvector components are equally de-localized, and in this case, the expectation of the IPR is $1/N$.
Fig.3(Right) shows the IPR versus the eigenvalue $\lambda_k, k=1,...,5$. The IPRs for $k=2,...,5$ approach $3/N$, which is the RMT expectation.
On the other hand, the IPR for $k=1$, the largest eigenvalue, is near $1/N$, which means that the eigenvector components are equally de-localized.
Fig.4 shows the time evolution of $IPR_1$ and the change of $IPR_1$, where the definition of the change of $IPR_1$ is the same as eq.(\ref{Change}).
The $IPR_1$ seems to decrease and approach $1/N$ at the three points found in the cross-correlation and the CRF,
although their signals are not very clear.
This observation on the $IPR_1$ indicates that when the market is volatile, the largest eigenvalue components are more equally de-localized,
which is different from the RMT expectation, and in such a case, the IPR approaches $1/N$.

\begin{figure}
\centering
\includegraphics[height=5.5cm]{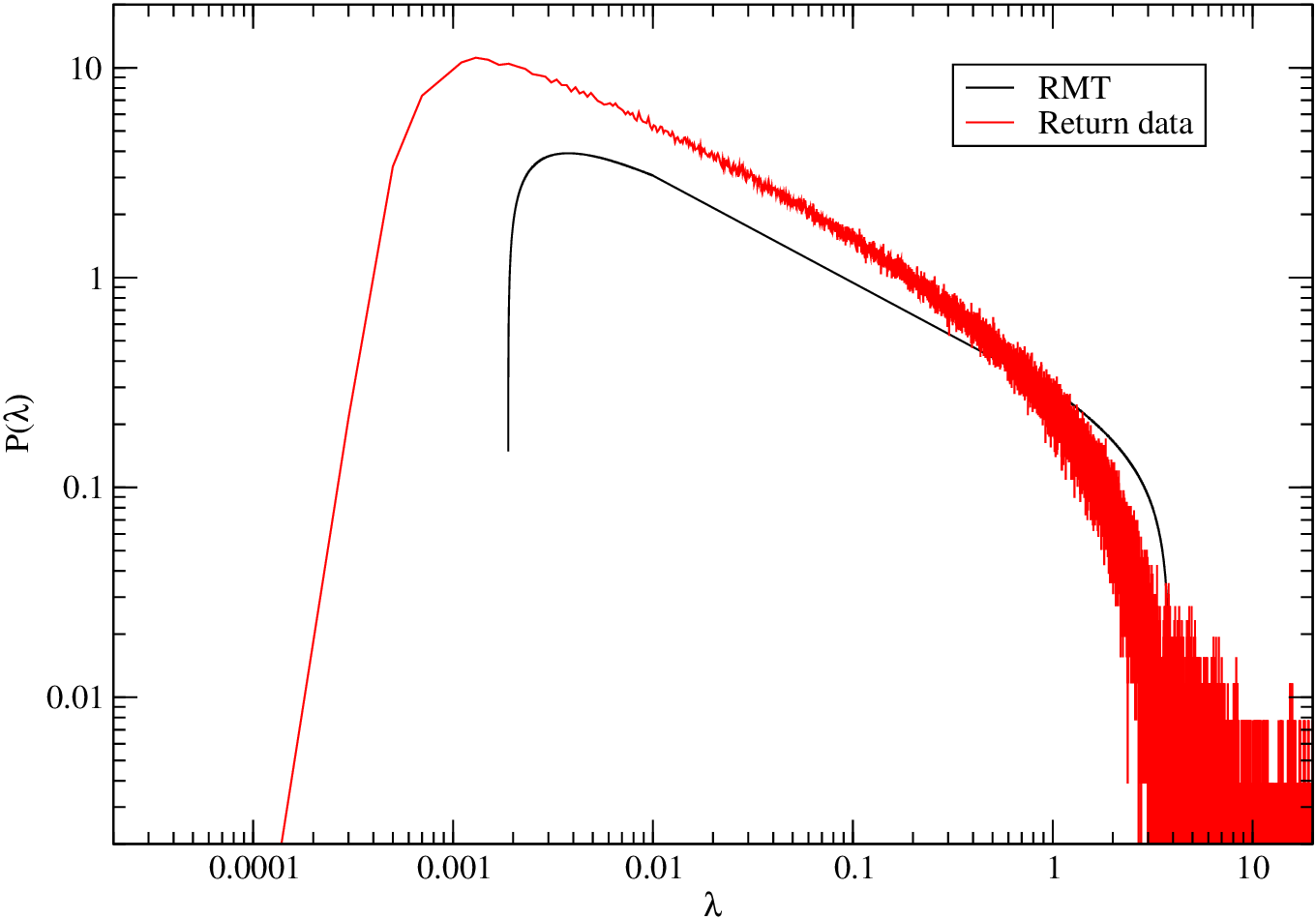}
\includegraphics[height=5.5cm]{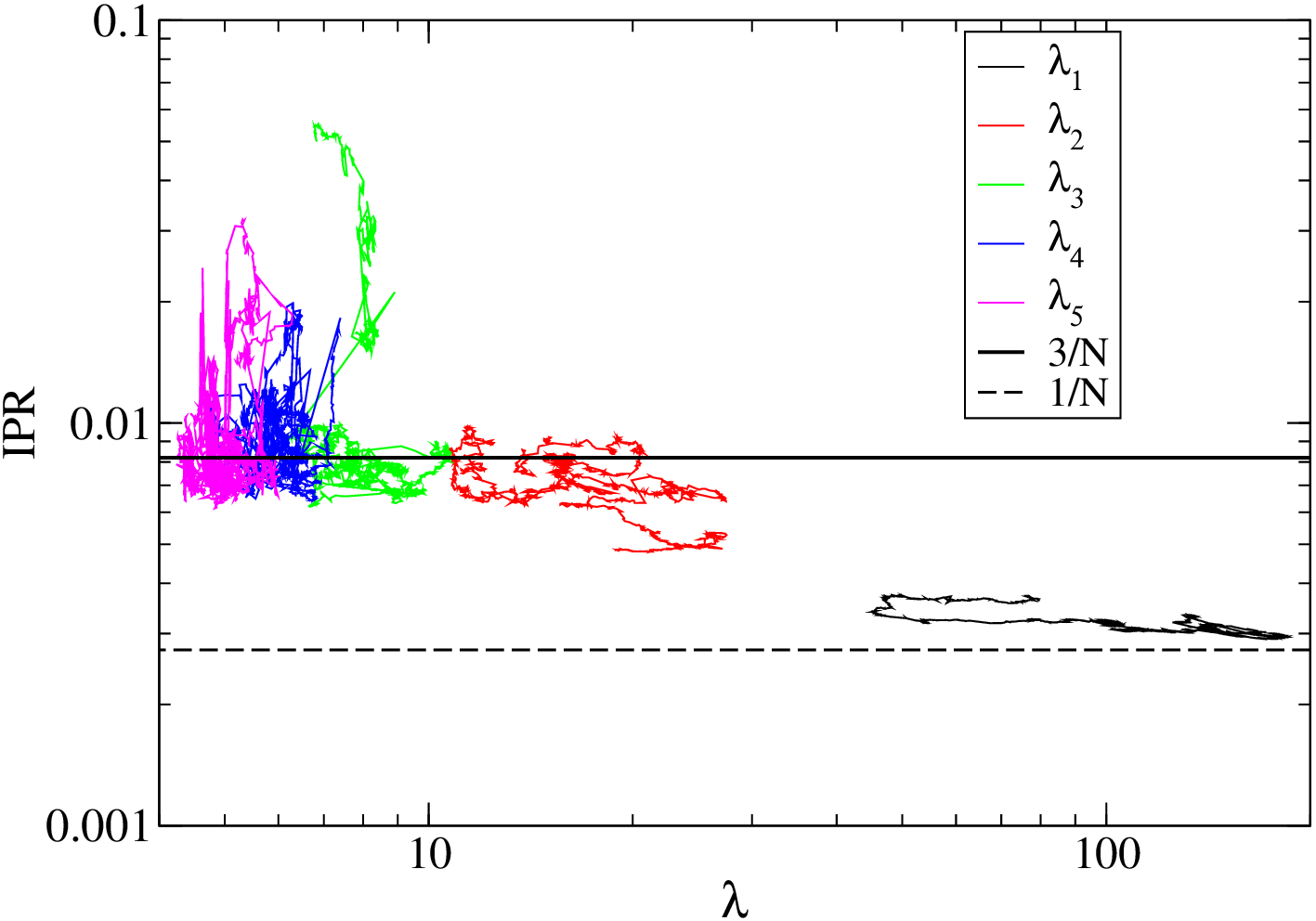}
\caption{
(Left) Eigenvalue distributions from the RMT and empirical return data.
(Right) IPR versus eigenvalues for $\lambda_1,...,\lambda_5$.
}
\vspace{-2mm}
\end{figure}

\begin{figure}
\centering
\includegraphics[height=7.0cm]{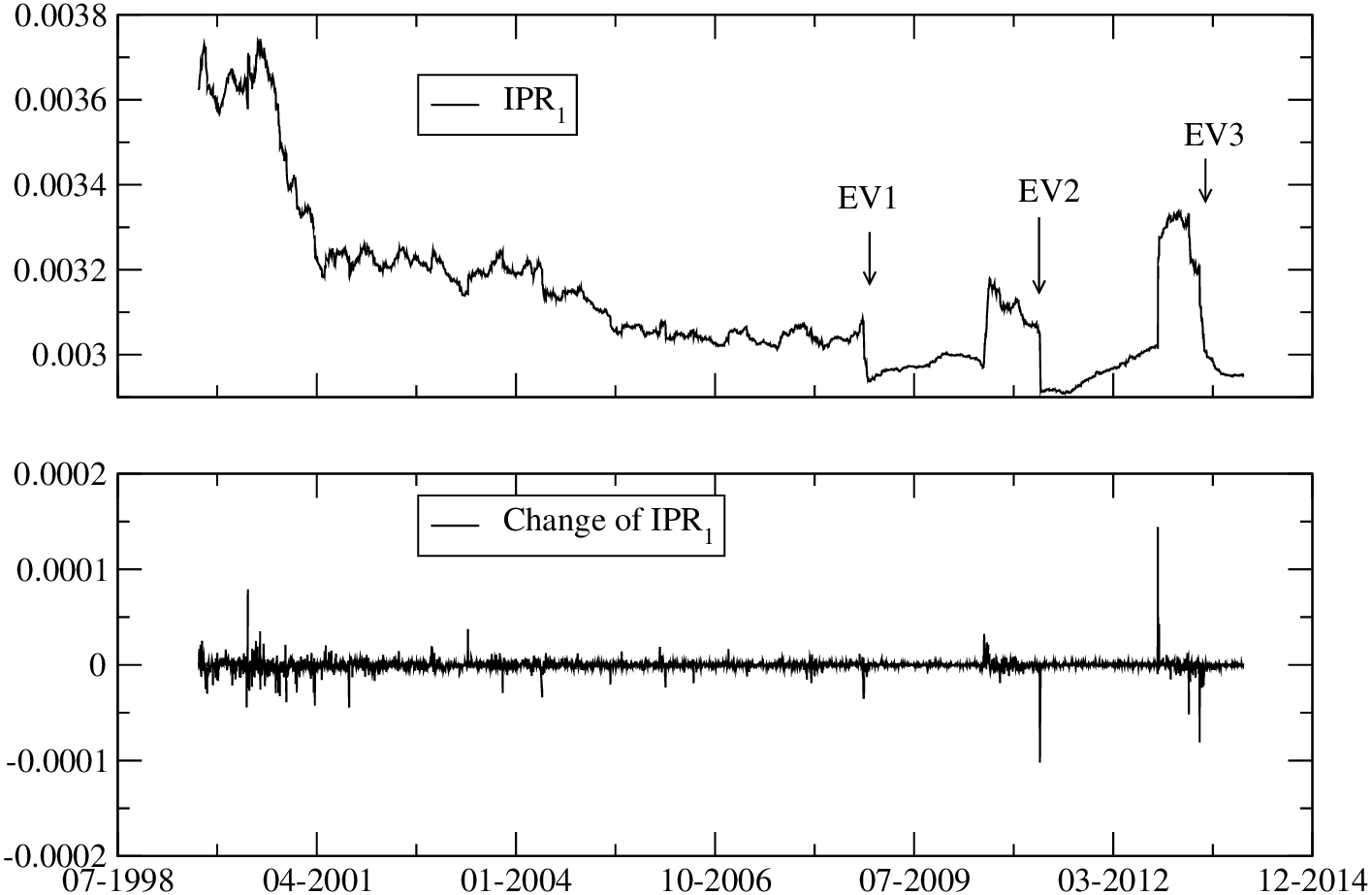}
\caption{
(Top) Time evolution of IPR and (Bottom) the change of IPR for $\lambda_1$.
EV1,...,EV3 correspond to the three events: the bankruptcy of Lehman Brothers, the Tohoku Region Pacific Coast Earthquake, and the FRB QE3 reduction observation,
respectively.
}
\vspace{-2mm}
\end{figure}

\section{Conclusions}
We have analyzed the cross-correlation matrices between 366 stocks traded on the Tokyo Stock Exchange
from January 5, 1998 to December 30, 2013.
We find that both the average off-diagonal elements of cross-correlation matrices and the cumulative risk fraction show
abrupt increases at three points that correspond to  three volatile stages of the Japanese stock market:
the bankruptcy of Lehman Brothers, the Tohoku Region Pacific Coast Earthquake, and the FRB QE3 reduction observation.
The change of the CRF also identifies these three points.
From comparison with the random matrix theory, we find that the empirical cross-correlation matrix differs from
the random matrix and, especially, the first eigenvector is more equally de-localized when the market is volatile.
The cross-correlation matrices contain relevant information on the financial market status.
By carefully analyzing the dynamical properties of the cross-correlations, we could monitor the risk that the financial markets confront.

\section*{Acknowledgement}
Numerical calculations in this work were carried out at the
Yukawa Institute Computer Facility
and the facilities of the Institute of Statistical Mathematics.
This work was supported by JSPS KAKENHI Grant Number 25330047.


\section*{References}
\begin{thebibliography}{99}
\bibitem{RMT1}
Plerou V, Gopikrishnan P, Rosenow B, Amaral L A N and Stanley H E 1999
\textit{Phys. Rev. Lett. }{\bf 83} 1471-1474

\bibitem{RMT2}
Laurent Laloux L, Cizeau P, Bouchaud J-P and Potters M 1999
\textit{Phys. Rev. Lett.} {\bf 83}  1467

\bibitem{RMT3}
Plerou V,  Gopikrishnan P, Rosenow B, Amaral L A N and Stanley H E  2000
\textit{Physica. A.} {\bf  287} 374-382

\bibitem{RMT4}
Plerou V,  Gopikrishnan P, Rosenow B, Amaral L A N,  Guhr T and Stanley H E  2002
\textit{Phys. Rev. E }{\bf 65} 066126

\bibitem{RMT5}
Utsugi A, Ino K and Oshikawa M 2004
\textit{Phys. Rev. E }{\bf 70} 026110

\bibitem{SR1}
Billio M, Getmansky M, Lo A W and Pelizzon L 2012
\textit{J. Financ Econ.} {\bf 104} 535

\bibitem{SR2}
Kritzman M, Li Y Z, Page S and Rigobon R 2011
\textit{J. Portf. Manag.} {\bf 37} 112
\bibitem{SR3}
Zheng Z, Podobnik B, Feng L and Li B 2012
\textit{Sci. Rep.} {\bf 2} 888
\bibitem{SR4}
Ren F and  Zhou W Z 2014
\textit{PLoS ONE} {\bf 9(5)}  e97711

\bibitem{W1}
Edelman A 1998
\textit{SIAMJ. Matrix Anal. Appl.} {\bf 9} 543
\bibitem{W2}
Sengupta A M and Mitra P P 1999
\textit{Phys. Rev. E} {\bf 60} 3389

\end{thebibliography}
\end{document}